# On the relation between the Feynman paradox and Aharonov-Bohm effects

**Scot McGregor, Ryan Hotovy, Adam Caprez, and Herman Batelaan**

Department of Physics and Astronomy, University of Nebraska—Lincoln, 208 Jorgensen Hall, Lincoln, Nebraska 68588-0299, USA
e-mail: hbatelaan2@unlnotes.unl.edu

**Abstract. The magnetic Aharonov-Bohm (A-B) effect occurs when a point charge interacts with a line of magnetic flux, while its dual, the Aharonov-Casher (A-C) effect, occurs when a magnetic moment interacts with a line of charge. For the two interacting parts of these physical systems, the equations of motion are discussed in this paper. The generally accepted claim is that both parts of these systems do not accelerate, while Boyer has claimed that both parts of these systems do accelerate. Using the Euler-Lagrange equations we predict that in the case of unconstrained motion only one part of each system accelerates, while momentum remains conserved. This prediction requires a time dependent electromagnetic momentum. For our analysis of unconstrained motion the A-B effects are then examples of the Feynman paradox. In the case of constrained motion, the Euler-Lagrange equations give no forces in agreement with the generally accepted analysis. The quantum mechanical A-B and A-C phase shifts are independent of the treatment of constraint. Nevertheless, experimental testing of the above ideas and further understanding of A-B effects which is central to both quantum mechanics and electromagnetism may be possible.**

## 1    Introduction

The question whether or not forces are present for physical systems that display the Aharonov-Bohm effect has been debated for decades. The general consensus is that there are no forces, which is considered to be a defining property of the famous effect. The best known version of the effect occurs when a current carrying solenoid (or more generally a magnetic flux) is enclosed by an electron interferometer. When the current is changed the consequence is that the observed electron fringes in the interferometer shift. Given that the solenoid is thought to produce no discernible magnetic (or electric) field external to its structure, and that is where the electron passes, there is no force on the electron. It is rare if not unique to encounter a response of a physical system without the presence of forces, which illuminates a part of the appeal of the A-B effect.

Central to A-B effects is the interaction between a magnetic moment and a charge.  This interaction is associated with a classical relativistic paradox [1]. Recently [2], Aharonov and Rohrlich stated that: "The paradox is crucial to clarifying the entirely quantum interactions of "fluxons" and charges – the generalized Aharonov-Bohm effect.."  The central problem to the paradox is the following. When a point charge moves in the vicinity of a tube that contains magnetic flux, the momentum in the electromagnetic field changes.  Outside of the flux tube there is no electric or magnetic field and the charge does not change its momentum.  The tube carries no net charge, may thus not experience a Lorentz force and appears not to change its momentum. These cursory observations would, if true, violate momentum conservation and give the appearance that the A-B effect is paradoxical in nature.

In this paper, we give a description of the magnetic A-B effect and its dual based on the Darwin Lagrangian. Our approach resolves the paradox, is consistent with all experiments to date, and can in principle be differentiated experimentally from previous theoretical approaches. We find that for constrained motion both parts of the physical system do not accelerate, consistent with the generally accepted prediction, however we also find that for unconstrained motion the magnetic part does accelerate and the charged part does not. The apparent violation of Newton's third law is typical for the "Feynman paradox."  The relation between the Feynman paradox and Aharonov-Bohm effects has to our knowledge not been pointed out before. Building on the Feynman paradox the difference between constrained and unconstrained motion is delineated. We argue that the appropriate description of physical systems, which are used for demonstration of A-B effects, is not known to be constrained or unconstrained.

Feynman explains a paradox in his famous Lectures where two particles interact in such a way that the momentum of one particle changes by a certain amount that is not the same as the momentum change of the other particle [3]. The specific scenario is that two charged particles are placed on the x-axis, with one charged particle moving initially along the x-axis, while the other moves along the y-axis. From the Lorentz force it is clear that the magnetic part of the force is not balanced (figure 1a). A relativistic treatment of this problem does not change this conclusion [4]. This is indeed an example where the interpretation of Newton's third law as conservation of mechanical momentum (as opposed to canonical momentum) breaks down.

In this work, a Lagrangian approach is chosen. The Lagrangian offers ways to conveniently impose constraints on the particle motion. A Hamiltonian can be obtained from it that can be compared to other approaches [5]. Finally, a path integral method can be used to obtain the quantum mechanical phase shifts that can be compared to the known A-B and A-C phase shifts. For the interaction of charged particles no Lagrangian exists that is manifestly invariant and obeys Lorentz symmetry [6] to all orders in $v/c$. The Darwin Lagrangian is the best known choice that is valid to $(v/c)^2$. This approximation will turn out to be sufficient to treat the Feynman paradox and the A-B and A-C problem in such a way that momentum is conserved, the equations of motion for both parts of the system are obtained and the method used for all systems is the same.

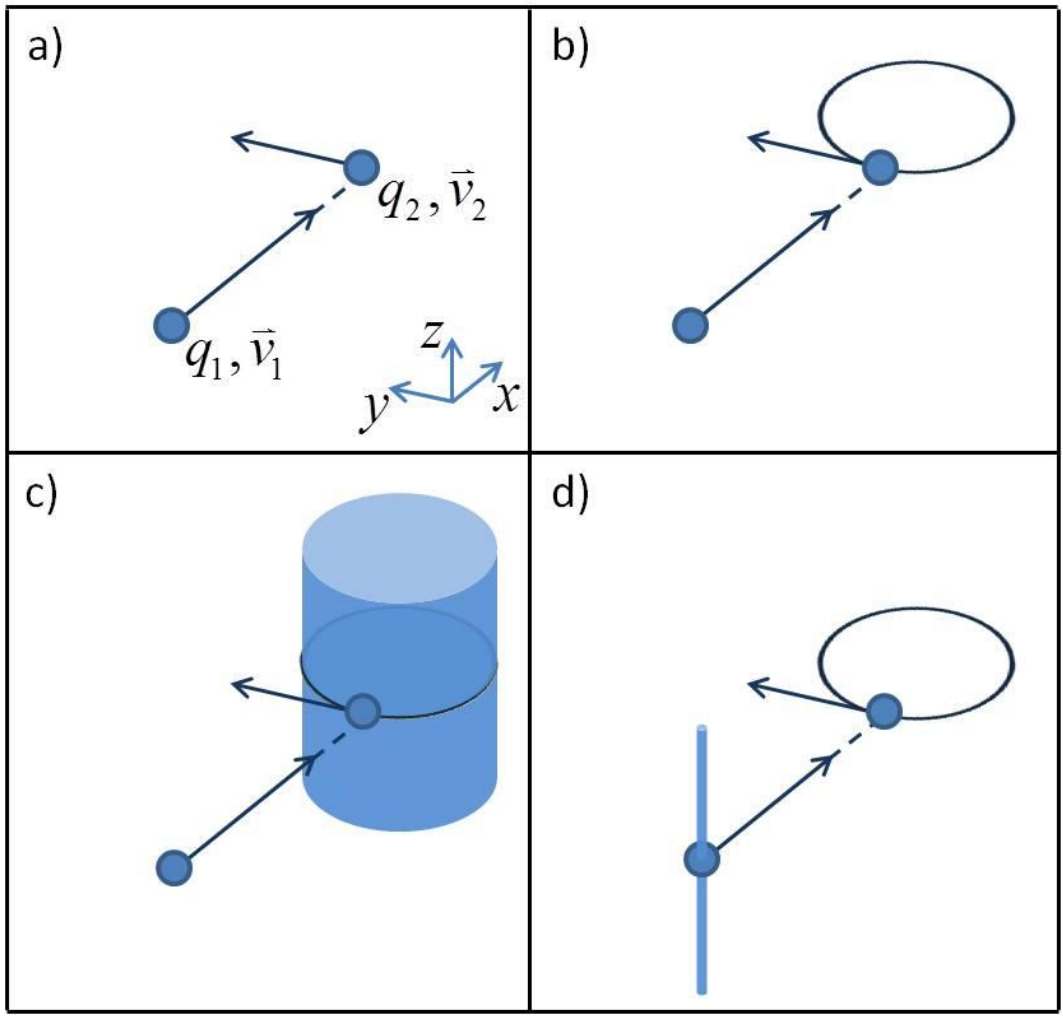

**Figure 1. Building physical systems. a) In the physical system presented in the Feynman paradox, particle 1 moves toward particle 2, and particle 2 moves with a velocity perpendicular to that of particle 1. The Lorentz Forces are not balanced in this case. b) The Mott-Schwinger system consists of a charged particle moving in the vicinity of a current loop [7, 8]. The current loop may be thought of as many circulating charge elements. Consequently this system bears a resemblance to the Feynman system. c) In the case of the Aharonov-Bohm effect, a charged particle is moving near a current carrying solenoid. Here the solenoid is depicted as constructed from current loops as they appear in the Mott-Schwinger system. d) The Aharonov-Casher system involves a charged wire and a current loop. Similar to the solenoid in the Aharonov-Bohm system, the charged wire is shown as constructed from charged particles as in the Mott-Schwinger system.**

## 2    Relativistic Classical Analysis

### 2.1    Preamble and assumptions: Building the physical systems

It is from the constituents of the physical system presented in the Feynman paradox (figure 1a) that the Mott-Schwinger system (figure 1b), the Aharonov-Bohm system (figure 1c), and the Aharonov-Casher system (figure 1d) can be constructed. The neutron in the Mott-Schwinger system can be modeled as a current loop. Such a loop may be thought of as many circulating charge elements. Thus, the transition from the Feynman paradox to the Mott-Schwinger system may be done by integration over the charges in the loop. Similarly a solenoid may be constructed via the addition of non-interacting current loops, and a charged wire constructed by addition of non-interacting point charges. Consequently, a transition from the Mott-Schwinger system to the Aharonov-Bohm or Aharonov-Casher systems may be done by integration of current loops or point charges, respectively.

In the construction phase the issue of constraints comes into play. The construction of the Mott-Schwinger system may be performed in two ways. Either the Lagrangian for the Feynman system can by integrated directly, or, alternatively, the forces resulting from the Lagrangian can be integrated. These two methods imply inherent assumptions regarding the freedom of the relative motion of the charges that constitute the current loop. If the forces resulting from the Lagrangian are integrated, the net force on the overall system, and thus the equation of motion of the current loop, is determined. Because the forces were computed without applying any restrictions to the relative motion, the charge elements are free to move independently (i.e. the motion of the charge elements is unconstrained). If, on the other hand, the Lagrangian is integrated directly, the Euler-Lagrange equations give the equation of motion for the current loop. The derivatives of the Euler-Lagrange equations are taken with respect to the position and velocity of the current loop. This method stipulates that the charge elements move relative to each other in such a way that the initial shape of the charge distribution is preserved and the loop merely undergoes translation (i.e. the motion of the charge elements is constrained).

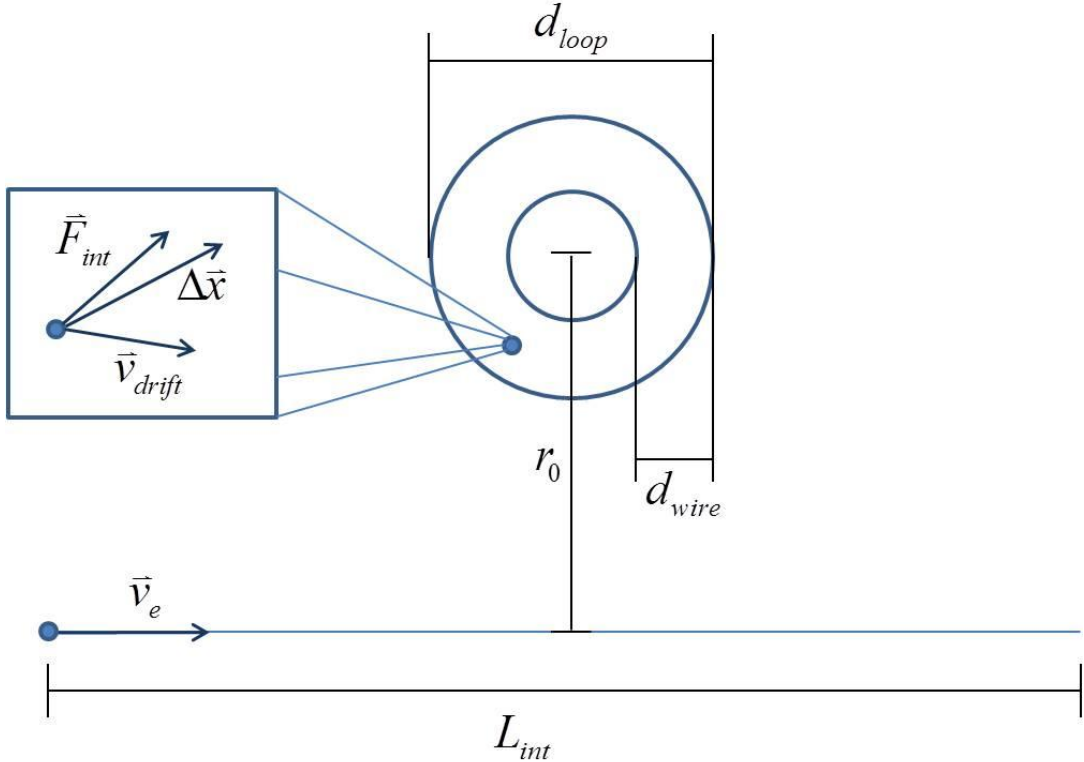

**Figure 2. Motion of a conduction electron. An electron in a current loop with diameter $d_{loop}$ and an electron passing at a distance $r_0$ interact via the Lorentz force. The electron in the loop experiences a force $F_{int}$. During the interaction time the electron in the loop moves a distance $\Delta x$. This movement is a combination of drift movement due to drift velocity $v_{drift}$ and the displacement due to the Lorentz force.**

It appears obvious that the motion of the conduction electrons in a solenoid should be treated as constrained. Simple estimates can be made to investigate this statement. Consider an electron passing a solenoid in a certain interaction time. During this time the motion of solenoidal conduction electrons can be investigated and their distance traveled can be compared to the solenoid wire thickness. If the distance traveled is much larger, then constraints are certainly important, while if the distance traveled is much shorter the roll that the constraints play is much less clear. Our argumentation hinges on the veracity of the latter and justifies the investigation of comparison of motion for unconstrained versus constrained systems. *We do not claim that the system is either*, but consider both fully unconstrained and constrained systems to be interesting limiting cases.

In A-B experiments such as the one by Mollenstedt and Bayh [9], the interaction time of an electron passing a solenoid at 40 keV is roughly 1 ps (see figure 2), assuming an interaction length of three times the loop diameter ($3 \times 36$ µm). The electron velocity has a drift velocity of $v_{drift} = I/nAq = 80$ µm/s, where $I$ is the current, $n$ is the number of atoms per unit volume of the wire, $A$ is the cross sectional area of the wire and $q$ is the charge of an electron. The electron has a far larger thermal component $v_{thermal} = \sqrt{2k_B T/m_e} = 9.5 \times 10^5$ m/s. The thermal drift displacement during the interaction time is $\Delta x_{thermal}$ = 87 nm, which is much smaller than the solenoid wire diameter of 5 µm. The displacement of electrons within the coil due to the magnetic field of the passing electron $\Delta x_{int}$ can also be approximately determined, by using the Lorentz force. The result is $\Delta x_{int} = 3.7 \times 10^{-20}$ m using the thermal velocity. Note that the inclusion of the effective electron mass of the Drude-Sommerfeld model has little effect on the estimates, as the effective mass of a conduction electron in tungsten is only 2-3 times that of a free electron [10]. The potential which restricts the charge to the wire may be thought of as having negligible curvature over such small distances. Additionally, the centripetal force required for the electrons in the solenoid to move in a circle with a drift velocity of 80 µm/s is on the order of $10^{-34}$ N, whereas the Lorentz force due to the passing electron charge is on the order of $10^{-32}$ N. It appears reasonable to at least consider the scenario of unconstrained motion.

Objections can be raised to these estimates. For example, the interaction time is much slower than the plasmonic response time of tungsten (0.44 fs) [11]. This motivates the inclusion of electron-electron interaction within the wire during the interaction time. An interesting attempt has been made to include such interactions and some constraints [12], that support the controversial idea that both parts of the A-B system experience a force. However, arguably [13], a recent experiment may rule out the presence of force on the passing electron [14]. To date, no detailed models have been analytically or numerically solved, which motivates the study of the simpler case of constrained and unconstrained motion.

For neutrons in the A-C system this type of estimate gives a completely different result. The neutron could be modeled as a current loop of radius $10^{-15}$ m. (This simplistic classical model ignores quantum mechanical addition of quark angular momentum and magnetic moment). In order for such a loop to generate a magnetic moment of $10^{-26}$ J/T, the constituent charges would circulate with a period on the order of $10^{-23}$ s. The interaction time in the experiment by Werner *et al.* [15] was on the order of $10^{-5}$ s thus the motion of the charged constituents of the neutron is constrained. For completeness it is still interesting to analyze the A-C system in terms of constrained and unconstrained motion as described above. Furthermore, A-C phase shift may be observable for other larger magnetic particles, for which the constraints are not clear.

A case has been made in favor of the effective presence of constraints on the basis of the following lemma: any finite stationary distribution of matter has zero total momentum [16]. The term "stationary" is defined by $\partial_0 T^{\mu\nu} = 0$, where $T^{\mu\nu}$ is the electromagnetic stress tensor. The assumption of a stationary distribution along with the conservation law $\partial_\mu T^{\mu\nu} = 0$ gives the result $\partial_j T^{j0} = 0$. Using the divergence theorem the total momentum may be written as a surface integral [17]

$$p^i = \frac{1}{c}\int T^{i0}d\tau = \frac{1}{c}\int \left[\partial_j\left(x_i T^{j0}\right) - x_i\partial_j T^{j0}\right]d\tau = \frac{1}{c}\oint x_i T^{j0}dS_j \; . \tag{1}$$

The assumption of a finite distribution of matter ensures that the elements of the stress tensor must fall off as $1/r^{4+\delta}$ $(\delta \geq 0)$. Consequently the above surface integral is zero, proving the lemma;

$$p^i = \frac{1}{c}\oint x_i T^{j0}dS_j = 0 \; . \tag{2}$$

The presence of electromagnetic momentum for a stationary charge-current distribution, taken together with the validity of the lemma, demands that there is another opposite and equal form of momentum. This "hidden momentum" results from internal motion of a stationary system. One text-book example is that of a current carrying loop of wire, bathed in a uniform external electric field [18] (figure 3). Relevant for our present discussion, the electric field could be thought of as arising from the presence of a distant point charge.

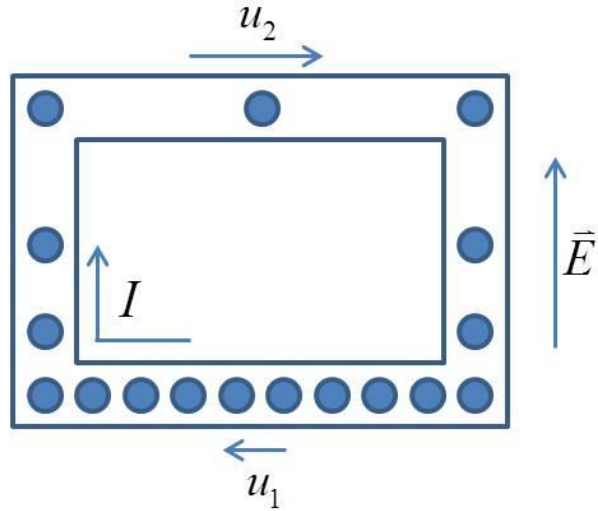

Figure 3: Hidden momentum. A conducting loop with current circulating clockwise is immersed in an external homogeneous electric field directed toward the top of the page. The electric field accelerates the charges moving toward the top of the loop and decelerates those moving toward the bottom of the loop. Consequently there is a non-zero net relativistic total linear momentum of the charges contained in the loop [18]. This is the "hidden momentum" and it exactly cancels the momentum in the electromagnetic field.

The applied electric field $\vec{E}$ gives rise to a change in velocity of the charges as they move along the vertical segments of the loop. Consequently, the velocity of the charges moving in the bottom segment, $u_1$, is smaller than the velocity in the top section, $u_2$. The result is that the charges in the loop carry a net relativistic mechanical momentum equal and opposite to the electromagnetic field momentum [18]. Proponents of using hidden momentum for analysis of the A-B effects, claim that in the case of dynamic systems for which equations of motion are being calculated, the hidden momentum has a direct effect on the equation of motion of the object in question. In the case of a current loop passing a charged wire (A-C system) the "hidden momentum" goes directly into the equation of motion so as to cancel the force on the loop. However, one should tread carefully when taking this approach considering that the lemma being applied requires a stationary system while the calculation of the equations of motion of a system requires the assumption of a non-stationary system. Such an analysis of the loop-wire system has been made with three different models of the current loop [16]: a gas of charged particles constrained to move inside a neutral tube, a gas of charged particles constrained to move inside a conducting tube, a charged (incompressible) fluid constrained to move inside a neutral tube. Although these analyses all predict zero forces, this is not a general property for unconstrained motion as shown by the counterexample given in our present analysis.

### 2.2 Unconstrained motion

In section 2.2.1 the force and the equations of motion for two interacting charged particles is derived from the Darwin Lagrangian for the Feynman problem (figure 1a). In the following two sections the force is integrated for charge and current distributions that are relevant for the Mott-Schwinger, and the A-B and A-C effects, respectively.

### 2.2.1 Equations of motion for two interacting charged particles using the Darwin Lagrangian

The Darwin Lagrangian [17] is given by

$$L = \frac{1}{2}m_1 v_1^2 + \frac{1}{2}m_2 v_2^2 - \frac{q_1 q_2}{r} + \frac{q_1 q_2}{2rc^2}\left[\vec{v}_1 \cdot \vec{v}_2 + \frac{(\vec{v}_1 \cdot \vec{r})(\vec{v}_2 \cdot \vec{r})}{r^2}\right], \tag{3}$$

where $\vec{r} = \vec{r}_1 - \vec{r}_2$. The vector potential and scalar potential for a moving charged particle are given by

$$\vec{A} = \frac{q}{2rc}\left[\vec{v} + \frac{\vec{r}(\vec{v} \cdot \vec{r})}{r^2}\right] \tag{4}$$

$$\varphi = \frac{q}{r}. \tag{5}$$

The Euler-Lagrangian equations of motion [19] are $\dfrac{d}{dt}\dfrac{\partial L}{\partial \vec{v}_1} = \dfrac{\partial L}{\partial \vec{r}_1}$ and $\dfrac{d}{dt}\dfrac{\partial L}{\partial \vec{v}_2} = \dfrac{\partial L}{\partial \vec{r}_2}$, where

$$\frac{d}{dt}\frac{\partial L}{\partial \vec{v}_1} = m_1\vec{a}_1 - \frac{q_1 q_2(\vec{r}\cdot\dot{\vec{r}})}{2c^2 r^3}\left[\vec{v}_2 + \frac{(\vec{v}_2\cdot\vec{r})\vec{r}}{r^2}\right] + \frac{q_1 q_2}{2c^2 r}\left\{\vec{a}_2 - \frac{2(\vec{v}_2\cdot\vec{r})(\vec{r}\cdot\dot{\vec{r}})\vec{r}}{r^4} + \frac{[(\vec{a}_2\cdot\vec{r})+(\vec{v}_2\cdot\dot{\vec{r}})]\vec{r}+(\vec{v}_2\cdot\vec{r})\dot{\vec{r}}}{r^2}\right\}$$

$$= m_1\vec{a}_1 + \frac{q_1 q_2}{2c^2 r}\left\{\vec{a}_2 - \frac{(\vec{r}\cdot\dot{\vec{r}})\vec{v}_2}{r^2} - \frac{3(\vec{v}_2\cdot\vec{r})(\vec{r}\cdot\dot{\vec{r}})\vec{r}}{r^4} + \frac{[(\vec{a}_2\cdot\vec{r})+(\vec{v}_2\cdot\dot{\vec{r}})]\vec{r}+(\vec{v}_2\cdot\vec{r})\dot{\vec{r}}}{r^2}\right\} \tag{6}$$

$$\frac{d}{dt}\frac{\partial L}{\partial \vec{v}_2} = m_2\vec{a}_2 + \frac{q_1 q_2}{2c^2 r}\left\{\vec{a}_1 - \frac{(\vec{r}\cdot\dot{\vec{r}})\vec{v}_1}{r^2} - \frac{3(\vec{v}_1\cdot\vec{r})(\vec{r}\cdot\dot{\vec{r}})\vec{r}}{r^4} + \frac{[(\vec{a}_1\cdot\vec{r})+(\vec{v}_1\cdot\dot{\vec{r}})]\vec{r}+(\vec{v}_1\cdot\vec{r})\dot{\vec{r}}}{r^2}\right\} \tag{7}$$

$$\frac{\partial L}{\partial \vec{r}_1} = \frac{q_1 q_2}{r^3}\vec{r} + \frac{q_1 q_2}{2c^2}\left[\frac{-(\vec{v}_1\cdot\vec{v}_2)\vec{r}}{r^3} - \frac{3(\vec{v}_1\cdot\vec{r})(\vec{v}_2\cdot\vec{r})\vec{r}}{r^5} + \frac{(\vec{v}_1\cdot\vec{r})\vec{v}_2+(\vec{v}_2\cdot\vec{r})\vec{v}_1}{r^3}\right] \tag{8}$$

$$\frac{\partial L}{\partial \vec{r}_2} = -\frac{q_1 q_2}{r^3}\vec{r} - \frac{q_1 q_2}{2c^2}\left[\frac{-(\vec{v}_1\cdot\vec{v}_2)\vec{r}}{r^3} - \frac{3(\vec{v}_1\cdot\vec{r})(\vec{v}_2\cdot\vec{r})\vec{r}}{r^5} + \frac{(\vec{v}_1\cdot\vec{r})\vec{v}_2+(\vec{v}_2\cdot\vec{r})\vec{v}_1}{r^3}\right]. \tag{9}$$

Taking the conditions which define the Feynman paradox (figure 4)

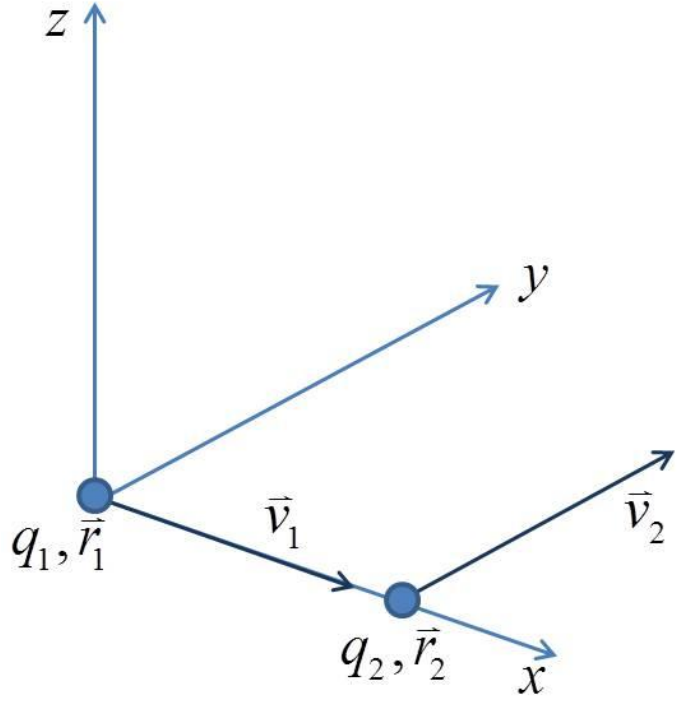

**Figure 4: The Feynman paradox. The coordinate system used for the analysis of the Feynman paradox (see text) is given.**

$$\vec{r}_1 = \vec{0}, \ \vec{r}_2 = r\hat{x}$$
$$\vec{v}_1 = v\hat{x}, \ \vec{v}_2 = v\hat{y}$$
$$q_1 = q_2, \ m_1 = m_2 \tag{10}$$
$$\vec{r} = -r\hat{x}, \ \dot{\vec{r}} = v(\hat{x} - \hat{y}), \ \hat{r} = -\hat{x}.$$

The equations of motion obtained for particle 1 are

$$a_{1x} = \frac{-\dfrac{q^2}{mr^2}\left[\left(1+\dfrac{v^2}{2c^2}\right)+\dfrac{q^2}{mc^2 r}\left(1-\dfrac{v^2}{c^2}\right)\right]}{1-\dfrac{1}{m^2}\left(\dfrac{q^2}{c^2 r}\right)^2} \approx -\frac{q^2}{mr^2}\left(1+\frac{v^2}{2c^2}\right) \tag{11}$$

$$a_{1y} = \frac{-\dfrac{q^2 v^2}{mc^2 r^2}}{1-\dfrac{1}{4m^2}\left(\dfrac{q^2}{c^2 r}\right)^2} \approx -\frac{q^2 v^2}{mc^2 r^2}, \tag{12}$$

and for particle 2

$$a_{2x} = \frac{\frac{q^2}{mr^2}\left[\left(1 - \frac{v^2}{c^2}\right) + \frac{q^2}{mc^2 r}\left(1 + \frac{v^2}{2c^2}\right)\right]}{1 - \frac{1}{m^2}\left(\frac{q^2}{c^2 r}\right)^2} \approx \frac{q^2}{mr^2}\left(1 - \frac{v^2}{c^2}\right) \tag{13}$$

$$a_{2y} = \frac{\frac{v^2}{2m^2 r}\left(\frac{q^2}{c^2 r}\right)^2}{1 - \frac{1}{4m^2}\left(\frac{q^2}{c^2 r}\right)^2} \approx 0. \tag{14}$$

The approximation in equations (11)-(14) is obtained by expansion to first order in $q^2/mc^2 r$ under the assumption that $q^2/mc^2 r \ll v^2/c^2$. This is valid if the paths of the charged particles are approximately straight. A small deflection implies that the potential energy of the particle is always less than the kinetic energy (i.e. $q^2/r < mv^2/2$). Alternatively, the relativistic equation of motion is given by the Lorentz force law

$$\vec{F} = q\left(\vec{E} + \frac{1}{c}\vec{v} \times \vec{B}\right). \tag{15}$$

Expanding the Lorentz force in this equation to second order in $v/c$ leads to the equations of motion:

$$a_{1x} = -\frac{\gamma q^2}{mr^2} \approx -\frac{q^2}{mr^2}\left(1 + \frac{v^2}{2c^2}\right) \tag{16}$$

$$a_{1y} = -\frac{\gamma q^2 v^2}{mc^2 r^2} \approx -\frac{q^2 v^2}{mc^2 r^2} \tag{17}$$

$$a_{2x} = \frac{q^2}{\gamma^2 mr^2} = \frac{q^2}{mr^2}\left(1 - \frac{v^2}{c^2}\right) \tag{18}$$

$$a_{2y} = 0, \tag{19}$$

which agree with the Darwin Lagrangian approach as well as Feynman's resolution of the paradox [4] in the non-relativistic limit. As Feynman points out, Newton's third law does not hold for mechanical momentum; however the consideration of the change of electromagnetic momentum ensures the conservation of total and canonical momentum. Note that the use of the Darwin Lagrangian is a superfluous step. We could have limited ourselves to the forces occurring in the relativistic equation of motion. However, for a consistent treatment of the unconstrained and constrained motion an identical starting point is favored. For unconstrained motion we can now proceed to integrate over the forces acting on the constituent particles of an extended body.

### 2.2.2    Charged particle and current loop

The forces in a system consisting of two interacting point charges have now been determined. A system of a point charge and a loop consisting of many mutually non-interacting point charges can now be constructed by direct integration over the forces. Consider a system consisting of a charged particle moving in the x direction in the vicinity of a current loop of radius $\varepsilon$ centered at the origin (figure 5).

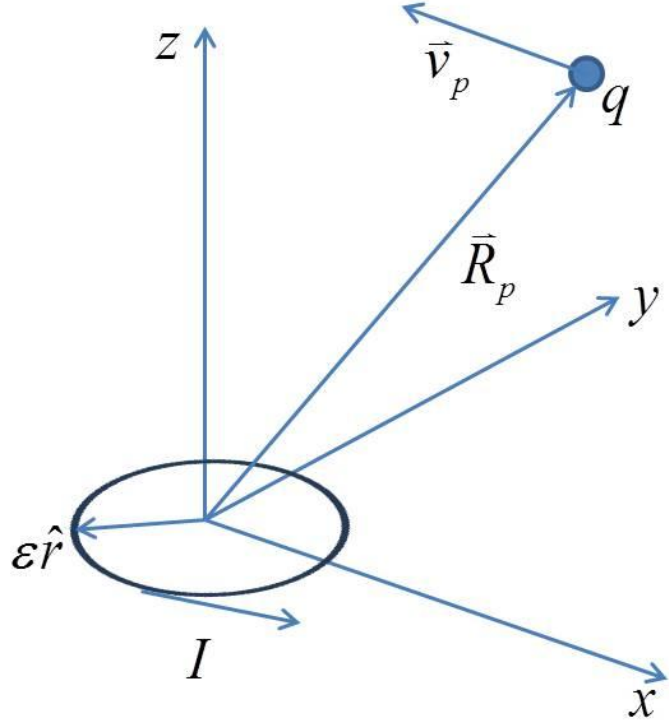

**Figure 5: The Mott-Schwinger system. The coordinate system for the analysis of a charged particle interacting with a current loop (see text) is given .**

$\vec{R}_q = \left( x_q, y_q, z_q \right)$ is the position of the charged particle relative to the center of the loop, $q$ is the charge, $\vec{v}_p$ is the velocity of the particle, and $I$ is the current. The force on the current loop due to the charged particle in the limit $\varepsilon \to 0$ is

$$\vec{F}_\mu = \frac{1}{c} \int \vec{J} \times \vec{B} d\tau = \frac{\mu q v_q}{c R_q^3} \left[ \frac{3 \left( x_q y_q \hat{x} + y_q^2 \hat{y} + y_q z_q \hat{z} \right)}{R_q^2} - \hat{y} \right]. \tag{20}$$

The force on the moving charge due to the current loop is

$$\vec{F}_q = \frac{q}{c} \vec{v}_q \times \vec{B} = \frac{\mu q v_q}{c R_q^3} \left[ \frac{3 \left( z_q^2 \hat{y} - y_q z_q \hat{z} \right)}{R_q^2} - \hat{y} \right], \tag{21}$$

where the magnetic moment is denoted by $\mu$. Note that the forces are not equal and opposite after integration and thus total mechanical momentum is not conserved similar to the Feynman paradox. The same procedure will now be followed for the Aharonov-Bohm and Aharonov-Casher systems (figure 1c and 1d).

### 2.2.3    Aharonov-Bohm and Aharonov-Casher systems

The forces involved in the Aharonov-Bohm (figure 1c) and Aharonov-Casher (figure 1d) systems can be determined by integration of the forces obtained for the loop/charge. For the A-B system the connection between the loop magnetic moment and the solenoid is made by substituting a differential magnetic moment element of the solenoid for the magnetic moment of the current loop:

$$\bar{\mu} \rightarrow \frac{c\Phi_B}{4\pi}\hat{z}dz_\mu,$$ (22)

where $\Phi_B$ is the magnetic flux in the solenoid. The charged particle is assumed to move in the x-direction. By integrating equation (21) the force on the charged particle is

$$\bar{F}_q = \int\limits_{solenoid} d\bar{F}_q = 0.$$ (23)

This is obvious given that the particle is propagating through a region where there are no electric or magnetic fields. By integrating equation (20) the force on the solenoid is

$$\bar{F}_s = \int\limits_{solenoid} d\bar{F}_\mu = \frac{q\phi_B v_q}{4\pi}\left\{\frac{2\left(x_q - x_s\right)\left(y_q - y_s\right)\hat{x} - \left[\left(x_q - x_s\right)^2 - \left(y_q - y_s\right)^2\right]\hat{y}}{\left[\left(x_q - x_s\right)^2 + \left(y_q - y_s\right)^2\right]^2}\right\}.$$ (24)

For the A-C system the connection between charged particle and the wire was made by substituting a differential charge element of the wire for the charge of the particle:

$$q \rightarrow \lambda dz_q.$$ (25)

By integrating equation (21) the force on the wire is

$$\bar{F}_w = \int\limits_{wire} d\bar{F}_q = 0$$ (26)

and by integrating equation (20) the force on the current loop is

$$\bar{F}_\mu = \int\limits_{wire} d\bar{F}_\mu = \frac{2\lambda\mu v_w}{c}\left\{\frac{2\left(x_w - x_\mu\right)\left(y_w - y_\mu\right)\hat{x} - \left[\left(x_w - x_\mu\right)^2 - \left(y_w - y_\mu\right)^2\right]\hat{y}}{\left[\left(x_w - x_\mu\right)^2 + \left(y_w - y_\mu\right)^2\right]^2}\right\}.$$ (27)

As stated in the introduction it is unreasonable to describe the motion of constituents of a neutron as unconstrained during the typical interaction times for A-C experiments. Moreover, the above simplistic reasoning foregoes the interesting physics that underlies the understanding of the neutron's magnetic moment as the sum of the magnetic moment of its parts and dynamics [20]. Nevertheless, for completeness in our present argument, the unconstrained model is considered in the context of the A-C physical system, and hopefully highlights the disparity in the nature of the solenoidal versus the neutron's magnetic moment. In this point of view Aharonov and Casher's realization that a neutron passing by a charged wire accumulates a phase shift that can be interpreted as the dual of the A-B effect is both beautiful and surprising.

In each of these systems one object feels a force while the other does not. This again is a system which exhibits the qualitative feature of the underlying Feynman system that total mechanical momentum is not conserved.

## 2.3    Constrained motion

In the following sections the integrated Lagrangian will be used to obtain the equations of motion for the Mott-Schwinger, A-B and A-C systems. The derivatives in the Euler Lagrange equation will be made with respect to coordinates that describe the motion of complete objects, such as the current loop in the Mott-Schwinger system. This constrains the motion of the charge elements in the loop to experience the same acceleration.

### 2.3.1    Integration of the Lagrangian

An alternative to the unconstrained method of analysis described above for the Mott-Schwinger system (figure 1b) is the approach of assuming that the charge elements within the loop are fixed relative to one another and must accelerate identically along with a coordinate defining the location of the loop. This can be done by two possible methods. By the first method, the vector potential of the moving charge, appropriate for the Darwin Lagrangian, is taken to determine the resulting magnetic field. The vector potential and magnetic field of the moving charge are

$$\vec{A}_q = \frac{q}{2rc}\left[\vec{v}_q + \frac{\vec{r}\left(\vec{v}_q \cdot \vec{r}\right)}{r^2}\right] \tag{28}$$

$$\vec{B}_q = \nabla \times \vec{A}_q = \frac{q}{c}\frac{\vec{v}_q \times \vec{r}}{r^3}. \tag{29}$$

The magnetic and electric fields are coupled to the magnetic dipole and relativistic electric dipole to obtain the Lagrangian

$$L = \frac{1}{2}m_q v_q^2 + \frac{1}{2}m_\mu v_\mu^2 + \vec{\mu}\cdot\vec{B} + \vec{d}\cdot\vec{E} = \frac{1}{2}m_q v_q^2 + \frac{1}{2}m_\mu v_\mu^2 + \frac{q}{c}\frac{\vec{\mu}\cdot\left[\vec{v}_q \times\left(\vec{r}_\mu - \vec{r}_q\right)\right]}{\left|\vec{r}_\mu - \vec{r}_q\right|^3} + \frac{q}{c}\frac{\left(\vec{v}_\mu \times \vec{\mu}\right)\cdot\left(\vec{r}_\mu - \vec{r}_q\right)}{\left|\vec{r}_\mu - \vec{r}_q\right|^3}$$

$$= \frac{1}{2}m_q v_q^2 + \frac{1}{2}m_\mu v_\mu^2 + \frac{q}{c}\frac{\left(\vec{v}_\mu - \vec{v}_q\right)\cdot\left[\vec{\mu}\times\left(\vec{r}_\mu - \vec{r}_q\right)\right]}{\left|\vec{r}_\mu - \vec{r}_q\right|^3}. \tag{30}$$

The second method is integration of the vector potential over the charges in the current loop. Integration of the vector potential (equation (23)) as it appears in the Darwin Lagrangian (equation (3)) for a current loop with no net charge gives

$$\vec{A}_\mu = \frac{\vec{\mu}\times\vec{r}}{r^3} \tag{31}$$

$$\varphi_\mu = \frac{1}{c}\vec{v}_\mu \cdot \vec{A}_\mu = \frac{\vec{v}_\mu \cdot \left(\vec{\mu}\times\vec{r}\right)}{cr^3}. \tag{32}$$

Coupling these potentials to the point charge gives the Lagrangian

$$L = \frac{1}{2}m_q v_q^2 + \frac{1}{2}m_\mu v_\mu^2 + \frac{q}{c}\,\vec{v}_q \cdot \vec{A}_\mu - q\varphi_\mu = \frac{1}{2}m_q v_q^2 + \frac{1}{2}m_\mu v_\mu^2 + \frac{q}{c}\frac{\vec{v}_q \cdot \left[\vec{\mu} \times \left(\vec{r}_q - \vec{r}_\mu\right)\right]}{\left|\vec{r}_q - \vec{r}_\mu\right|^3} - \frac{q}{c}\frac{\vec{v}_\mu \cdot \left[\vec{\mu} \times \left(\vec{r}_q - \vec{r}_\mu\right)\right]}{\left|\vec{r}_q - \vec{r}_\mu\right|^3}$$

$$= \frac{1}{2}m_q v_q^2 + \frac{1}{2}m_\mu v_\mu^2 + \frac{q}{c}\frac{\left(\vec{v}_\mu - \vec{v}_q\right)\cdot\left[\vec{\mu}\times\left(\vec{r}_\mu - \vec{r}_q\right)\right]}{\left|\vec{r}_\mu - \vec{r}_q\right|^3}. \tag{33}$$

These two methods give the same result due to the symmetry under permutation of particles of the Darwin Lagrangian and therefore only one should be taken for the computation of the equations of motion to avoid double counting. Applying the Euler-Lagrange equations gives

$$\frac{d}{dt}\frac{\partial L}{\partial \vec{v}} - \frac{\partial L}{\partial \vec{r}} = 0 \tag{34}$$

$$m_\mu \vec{a}_\mu = -\frac{q}{c}\left\{\frac{\left(\vec{v}_\mu - \vec{v}_q\right)\times\vec{\mu}}{\left|\vec{r}_\mu - \vec{r}_q\right|^3} + \frac{3\left[\left(\vec{r}_\mu - \vec{r}_q\right)\cdot\vec{\mu}\right]\left[\left(\vec{r}_\mu - \vec{r}_q\right)\times\left(\vec{v}_\mu - \vec{v}_q\right)\right]}{\left|\vec{r}_\mu - \vec{r}_q\right|^5}\right\} \tag{35}$$

$$m_q \vec{a}_q = \frac{q}{c}\left\{\frac{\left(\vec{v}_\mu - \vec{v}_q\right)\times\vec{\mu}}{\left|\vec{r}_\mu - \vec{r}_q\right|^3} + \frac{3\left[\left(\vec{r}_\mu - \vec{r}_q\right)\cdot\vec{\mu}\right]\left[\left(\vec{r}_\mu - \vec{r}_q\right)\times\left(\vec{v}_\mu - \vec{v}_q\right)\right]}{\left|\vec{r}_\mu - \vec{r}_q\right|^5}\right\}. \tag{36}$$

These forces are equal in magnitude and opposite in direction and thus conserve total mechanical momentum. Therefore this cannot be characterized as a Feynman type paradox.

The forces acting on the individual components of the A-B (figure 1c) and A-C (figure 1d) systems can be determined by integrating the Mott-Schwinger Lagrangian (equation (30) or (33)). The Lagrangian obtained for the A-B system is

$$L = \frac{1}{2}m_s v_s^2 + \frac{1}{2}m_q v_q^2 + \frac{q\Phi_B}{2\pi}\frac{\left(\vec{v}_q - \vec{v}_s\right)\cdot\left[\hat{z}\times\left(\vec{r}_q - \vec{r}_s\right)\right]}{\left(x_q - x_s\right)^2 + \left(y_q - y_s\right)^2} \tag{37}$$

$$= \frac{1}{2}m_q v_q^2 + \frac{1}{2}m_s v_s^2 + \frac{q}{c}\left(\vec{v}_q - \vec{v}_s\right)\cdot\vec{A}_s. \tag{38}$$

Likewise, the Lagrangian obtained for the A-C system is

$$L = \frac{1}{2}m_\mu v_\mu^2 + \frac{1}{2}m_w v_w^2 + \frac{2\lambda}{c}\frac{\left(\vec{v}_w - \vec{v}_\mu\right)\cdot\left[\vec{\mu}\times\left(\vec{r}_w - \vec{r}_\mu\right)\right]}{\left(x_w - x_\mu\right)^2 + \left(y_w - y_\mu\right)^2} \tag{39}$$

$$= \frac{1}{2}m_\mu v_\mu^2 + \frac{1}{2}m_w v_w^2 + \frac{1}{c}\left(\vec{v}_\mu - \vec{v}_w\right)\cdot\left(\vec{\mu}\times\vec{E}_w\right). \tag{40}$$

In both cases application of the Euler-Lagrange equations of motion gives zero force acting on both elements of both the A-B and A-C systems.

The predictions for the unconstrained motion are very different from the predictions of the constrained motion (the latter coinciding with generally accepted one). Can these two methods be distinguished by comparing their predicted phase shifts to the experimentally measured phase shifts?

## 3    Quantum mechanical phase shifts

### 3.1    Constrained

To compute the quantum mechanical phase shift for the charged particle and the neutron in the A-B and A-C effects, respectively, a closed loop path integral over time is taken for the Lagrangian described for constrained motion. The phase for the constrained case is the generally accepted one and only a brief summary is given in this section. In these calculations the charged wire and the solenoid are taken to be stationary ( $v_w = v_s = 0$ ). Using the Lagrangian given by equation (38) the A-B phase is

$$\varphi_{AB} = \frac{1}{\hbar} \oint \left( \frac{1}{2} m_q v_q^2 + \frac{q}{c} \vec{v}_q \cdot \vec{A}_s \right) dt = \frac{q\Phi_B}{\hbar c} \,, \tag{41}$$

which has been experimentally verified [9, 21-23]. Using the Lagrangian given by equation (40) the A-C phase is

$$\varphi_{AC} = \frac{1}{\hbar} \oint \left( \frac{1}{2} m_\mu v_\mu^2 + \frac{1}{c} \vec{v}_\mu \cdot \left( \vec{\mu} \times \vec{E}_w \right) \right) dt = \frac{4\pi\lambda\mu}{\hbar c} . \tag{42}$$

In either case the first term in the Lagrangian, ( $mv^2/2$ ), does not contribute to the phase. There is no force acting on the charged particle or the neutron and the effects are true A-B effects. An experimental test of the Aharonov-Casher effect by Werner *et al.* [5] is in agreement with the standard quantum mechanical prediction, where the experimental to theoretical ratio is given by $\varphi_{AC}^E / \varphi_{AC}^T = 1.46 \pm 0.35$ .

### 3.2    Unconstrained

In the path integral formulation [24] the wavefunction is propagated with the kernel, $K(b,a) = \exp\left( \frac{i}{\hbar} \int_{t_a}^{t_b} L \, dt \right)$ , where $L$ is the classical Lagrangian.   For a free particle the Kernel is $\exp\left( \frac{i}{\hbar} \int \vec{p} \cdot d\vec{r} \right)$ , where $p = mv$ . Formally, the initial wave function should now be constructed and propagated. However, for the purpose of understanding the measured phase shift in an interferometer it is customary to consider the effect on plane waves. In this case the phase shift is given by $\frac{1}{\hbar} \int_{t_a}^{t_b} L \, dt = \frac{1}{\hbar} \int_{t_a}^{t_b} (p\dot{x} - H) dt$ , where $p$ is the canonical momentum $p = mv + qA$ . In the case that the Hamiltonian is time independent the phase shift becomes $\frac{1}{\hbar} \int_{x_a}^{x_b} \vec{p} \cdot d\vec{x}$ [25].

For unconstrained motion in the case of the A-B effect the phase may therefore be written as follows

$$\varphi_{total} = \frac{1}{\hbar} \int \vec{p} \cdot d\vec{x} = \frac{1}{\hbar} \int \left( m\vec{v} + \sum q\vec{A}_j \right) \cdot d\vec{x}$$

$$\tag{43}$$

$$= \frac{1}{\hbar} \int \left( m\vec{v} + q\sum \vec{A}_j \right) \cdot d\vec{x} = \frac{1}{\hbar} \int \left( m\vec{v} + q\vec{A}_s \right) \cdot d\vec{x}$$

where $\vec{A}_s$ is the vector potential generated by the solenoid and $\vec{A}_j$ is the vector potential generated by the charges that constitute the solenoid. This is identical to the phase integral for the A-B effect in the case of constrained motion.

In the case of the A-C effect considering unconstrained motion as argued above is unreasonable. However, the existence of a larger particle with a magnetic moment cannot be excluded. Such a particle may have constituents that are best described by unconstrained motion. In our model, there are different forces acting on such constituents. How is the path integral phase shift defined for a composite object if the constituents experience different forces? The physical picture is that if the interaction does not lead to a change of the internal quantum states then the two arms of the interferometer remain indistinguishable. The measured phase shift reflects only the effect in the center of mass coordinate or external quantum state. If the internal quantum states do change then the contrast of the interferometer may be reduced. The initial wavefunction for an unconstrained composite particle with N mutually non-interacting constituents can be written as a product state of plane waves, $\psi_C = \prod_{j=1}^{N} \exp\left( i\, \vec{p}_j \cdot \vec{R}_j / \hbar \right)$. The phase accumulated by each plane wave along a path is $\varphi = \frac{1}{\hbar} \int \vec{p} \cdot d\vec{x}$ and thus the phase of the composite wavefunction $\psi_C$ picks up an overall phase factor of $\exp\left( \frac{i}{\hbar} \sum \int \vec{p}_j \cdot d\vec{x} \right)$. This phase factor may be rewritten in terms of the total force, $\vec{F}_{total}$, on the current loop as computed in section 2.2.3.,

$$\begin{aligned}
\varphi_{total} &= \frac{1}{\hbar} \sum \int \vec{p}_j \cdot d\vec{x} = \frac{1}{\hbar} \int \left( \sum \vec{p}_j \right) \cdot d\vec{x} = \frac{1}{\hbar} \int \left[ \sum \left( \vec{p}_{0j} + \int \vec{F}_j dt \right) \right] \cdot d\vec{x} \\
&= \frac{1}{\hbar} \int \left( \sum \vec{p}_{0j} \right) \cdot d\vec{x} + \frac{1}{\hbar} \int \left[ \int \left( \sum \vec{F}_j \right) dt \right] \cdot d\vec{x} = \frac{1}{\hbar} \int \left( \sum \vec{p}_{0j} \right) \cdot d\vec{x} + \frac{1}{\hbar} \int \left[ \int \vec{F}_{total} dt \right] \cdot d\vec{x} .
\end{aligned} \tag{44}$$

Note that the composite particle has no charge and the $qA$ term does not contribute to the phase. Integration of the total force (equation (27)) along a straight path gives the total phase

$$\varphi_{total} = \frac{1}{\hbar} \int \left( \sum \vec{p}_{0j} \right) \cdot d\vec{x} + \frac{2\pi\lambda\mu}{\hbar c} sign\left( y_\mu - y_w \right). \tag{45}$$

The difference in phase between the two paths is $\Delta\varphi_{total} = \frac{4\pi\lambda\mu}{\hbar c}$, which is the appropriate AC phase shift. Thus the constrained and unconstrained method cannot be distinguished by inspecting the phase.

## 4    Comparison to previous analyses.

### 4.1    Hidden momentum

The approach taken by Vaidman [16] as applied to the A-C system is one in which internal motion of the system manifest itself in "hidden momentum" which affects the motion of the neutron. The time derivative of this "hidden momentum" or the hidden force, as one may refer to it, is applied directly to the equation of motion

$$m\vec{a} = \frac{d\vec{p}}{dt} - \frac{d\vec{p}_{hid}}{dt} \; . \tag{46}$$

As mentioned above the justification for the use of the hidden momentum comes from a lemma that states that for stationary and finite current and charge distributions the total momentum is zero. A non-zero value of the electromagnetic field momentum than implies the presence of a hidden momentum of equal magnitude and opposite in direction:

$$\vec{p}_{hid} = -\frac{1}{c^2} \int \varphi \vec{J} d\tau = -\frac{1}{4\pi c} \int \vec{E} \times \vec{B} d\tau = -\vec{p}_{em} , \tag{47}$$

where $\varphi$ is the electrostatic potential of the charged wire and $\vec{J}$ is the current density of the loop. The electric potential and current density result in an electric field $\vec{E}$ and magnetic field $\vec{B}$, respectively. Thus the equation of motion explicitly depends on the change of the electromagnetic field momentum,

$$m\vec{a} = \frac{d\vec{p}}{dt} + \frac{d}{dt} \left[ \frac{1}{4\pi c} \int \vec{E} \times \vec{B} d\tau \right]. \tag{48}$$

The equation of motion for a current loop in the Aharonov-Casher system (figure 1d) determined by direct application of this method is

$$\begin{aligned}
m\vec{a} = \frac{d\vec{p}}{dt} - \frac{d\vec{p}_{hid}}{dt} &= \nabla\left(\vec{\mu} \cdot \vec{B}\right) - \frac{1}{c}\frac{d}{dt}\left(\vec{\mu} \times \vec{E}\right) \\
&= -\frac{1}{c} \nabla\left[\vec{\mu} \cdot \left(\vec{v} \times \vec{E}\right)\right] - \frac{1}{c}\frac{d}{dt}\left(\vec{\mu} \times \vec{E}\right) \\
&= -\frac{1}{c}\left[\left(\vec{\mu} \cdot \nabla\right)\left(\vec{v} \times \vec{E}\right) - \left(\vec{v} \cdot \nabla\right)\left(\vec{\mu} \times \vec{E}\right)\right] - \frac{1}{c}\left(\vec{v} \cdot \nabla\right)\left(\vec{\mu} \times \vec{E}\right) \\
&= -\frac{1}{c}\left(\vec{\mu} \cdot \nabla\right)\left(\vec{v} \times \vec{E}\right).
\end{aligned} \tag{49}$$

This acceleration is zero for the geometry of the Aharonov-Casher effect. Thus the force on both objects in the Aharonov-Casher system is zero, by this method.

However, for the Feynman paradox the equations of motion do not depend on the change of the electromagnetic field momentum. The inclusion of electromagnetic field momentum solves the paradox by offering a third physical entity that carries a changing momentum [4], while the forces on both objects are not zero, contrasting the Vaidman analysis of the Aharonov-Casher system. Why is there a difference between the two analyses? The reason is that the Feynman paradox concerns a physical system that is not a stationary charge distribution and the Lemma does not hold. The question for the A-C system is than if it is well represented by a stationary charge and current distribution. Clearly, the neutron passes by the charged wire and formally, the A-C system is not represented by a stationary distribution. The result that our constrained description gives is the same as the Vaidman approach, while it is interesting to consider the unconstrained result in relation to the Feynman paradox.

### 4.2 Newton's third law

The approach taken by Boyer is documented in a series of papers that extend over several decades [12, 13, 26, 27], and argue that the Aharonov-Bohm effects are accompanied by a force. This point of view conflicts the generally accepted interpretation of the A-B effect. We will limit ourselves to comment on two of the more recent papers in this series. Boyer considers a charged particle passing by a solenoid (represented by a line of magnetic dipoles) and calculates the Lorentz force on the solenoid [12]. This force is the same as that given in section 2.2.3 (equation (24)) and Boyer's work motivated that part of our calculation. Boyer continues his argument by invoking Newton's third law and noticing that the back-acting force on the electron causes a displacement that through a semi-classical argument gives exactly the Aharonov-Bohm phase shift. It is remarkable that such an argument can be given that provides exactly the necessary force, in view of the observation that an unperturbed solenoid has no external electromagnetic fields. The argument hinges on three assumptions. First the force on the solenoid is the total force that acts on the solenoid, second Newton's third law holds, and third the semi-classical approximation is valid. Our work shows that the total force on the solenoid depends on the presence or absence of constraints. Additionally, Feynman's paradox illustrates that Newton's third law is not generally valid. (Boyer argues in another paper in 2002 that the electromagnetic momentum is conserved during the interaction [12]). Finally it is interesting to note that Boyer's force is dispersionless, implying that the group velocity of a wavepacket in a semiclassical approximation does not change. All these issues are interesting in their own right, and warrant further discussion.

In a paper that comments on our experimental demonstration of the absence of force for a charged particle passing a solenoid [13], Boyer argues that charged particles in a solenoid that mutually interact and experience friction can provide a back-action on the passing particle. This line of reasoning considers a model that is more complex than the ones considered previously and in our current paper, because mutual interaction between the constituents of magnetic dipoles are excluded.

### 4.3 Hamiltonian approach

An analysis based on a Hamiltonian approach by Werner and Klein [5] has been done to determine the force on the neutron in the Aharonov-Casher system (figure 1d). The Hamiltonian used was

$$H = \frac{p^2}{2m} - \frac{1}{mc}\,\vec{\mu}\cdot\left(\vec{E}\times\vec{p}\right). \tag{50}$$

A direct application of Hamilton's equations of motion gives

$$\dot{\vec{r}} = \frac{\partial H}{\partial \vec{p}} \tag{51}$$

$$\dot{\vec{p}} = -\frac{\partial H}{\partial \vec{r}} \tag{52}$$

$$m\ddot{\vec{r}} = -\frac{1}{c}\left(\vec{\mu}\cdot\nabla\right)\left(\vec{v}\times\vec{E}\right). \tag{53}$$

In the Aharonov-Casher geometry the electric field has no spatial dependence in the direction of the magnetic moment, therefore, the force on the neutron is zero, by the above prescription. Note that this approach does not describe a closed system as it is a single particle Hamiltonian. Because this approach is that of an open system it does not address conservation of momentum. Thus, the criterium that total momentum must be conserved cannot be applied to this approach as a test of the validity of the Hamiltonian. Furthermore, this Hamiltonian is equivalent to our Lagrangian (equation (40)) for a stationary wire. Using the vector identity $\left(a\times b\right)\cdot c = a\cdot\left(b\times c\right)$ the equivalence is found:

$$L = \frac{1}{2}mv^2 + \vec{d} \cdot \vec{E} = \frac{1}{2}mv^2 + \frac{1}{c}\left(\vec{v} \times \vec{\mu}\right) \cdot \vec{E} \tag{54}$$

$$\vec{p} = \frac{\partial L}{\partial \vec{v}} = m\vec{v} + \frac{1}{c}\vec{\mu} \times \vec{E} \tag{55}$$

$$H = \vec{p} \cdot \vec{v} - L = \frac{1}{2m}\left(\vec{p} - \frac{1}{c}\vec{\mu} \times \vec{E}\right)^2 \approx \frac{p^2}{2m} - \frac{1}{mc}\vec{p} \cdot \left(\vec{\mu} \times \vec{E}\right). \tag{56}$$

This Hamiltonian can thus be classified as describing a constrained system as described in section 2.3.1.

### 4.4    Aharonov and Rohrlich

In their 2005 book, "Quantum Paradoxes: Quantum Theory for the Perplexed", Aharonov and Rohrlich discuss various momentum terms that can make up for the changing momentum in the electromagnetic field and ultimately conserve momentum. The missing momentum is stated to be the relativistic momentum of the charged particles which give rise to the magnetic flux. The contribution of the Lorentz force to the momentum conservation is ignored. The statement that "We move it [passing particle] as slowly as we like, so that the charge scarcely induces a magnetic field..." does not address this issue. Although the magnetic field and thus the Lorentz force scale linearly with velocity, the momentum exchange is independent of velocity as the interaction time scale inversely with velocity. In this paper it is shown that (in the unconstrained description) the change of momentum due to the Lorentz force is identical in magnitude to the change of momentum in the electromagnetic field.

## 5    Conclusion

The relation between the Feynman paradox and the AB-effects is that an unconstrained treatment of the AB-effects share with the Feynman paradox the property that momentum is stored in the electromagnetic field during the interaction, and consequently that the forces on the two interacting mechanical parts of the system are not balanced. This implies that one part of the system experiences a force, which is a prediction that is in stark contrast with the usual understanding of AB-effects. In the constrained description the AB-effects are very different than the Feynman paradox. In this description, the usual prediction is made that both mechanical parts do not experience a force. Both of these scenarios are limited to the case that constituents that make up the magnetic moment are assumed not to interact. Given the limited theoretical scope of the theoretical claims, experiments are important. However, as we will indicate now, there are very few options within reach of present technology.

An experiment to test for the force on an electron in the Aharonov-Bohm system (figure 1c) has been conducted by our group (see Caprez *et al.*). In that experiment a time delay was measured for an electron passing between two solenoids [14]. The time required for the electron to pass from source to detector was found to be independent of the magnetic flux contained in the solenoids and thus it appears that the Aharonov-Bohm phase shift cannot be explained by a classical force on the electron. However, it has been pointed out that in this case a macroscopic solenoid was used and the qualitative characteristics of the system, such as whether or not there is a measurable delay, potentially depend on the size of the solenoid [13]. For larger solenoids the interaction time is greater than the plasma oscillation period. This is the case for all experimental tests of the A-B effect so far, and as such the force experiment and phase experiments are performed in the same regime. The issue considered in this paper is a different one. The above experiment does not discriminate between the unconstrained and constrained description.

For the Aharonov-Bohm system, an experiment to detect the predicted force on the solenoid (as predicted by the unconstrained model) appears impossible given the necessity to detect the force of a single electron on a macroscopic object.

Although experiments have been done to show the Aharonov-Casher phase shift, no experiments have tested for the presence of a force on the neutron. However, for the molecule Thallium Fluoride the phase shift was shown to be independent of velocity [28] which is a feature associated with the dispersionless nature of the A-B effect and signals the absence of force [29]. The interaction between the applied electric field and the magnetic moment of the fluoride nucleus was responsible for the phase shift. Given the small size of a nucleus, or even an atom or molecule that may be used in such type of experiments, the circulation time for constituent charges that produce the magnetic moment is much less than the interaction time. It is likely then that the system must be modeled by constrained motion. Consequently, our present analysis would predict that there is, in fact, no force acting on the interfering particle, consistent with the Thallium Fluoride experiment.

Similarly, due to the small size of the neutron, the Mott-Schwinger effect for neutron scattering of nuclei is not a physical system that can provide an interesting test between the constrained and unconstrained description. On the other hand if the magnetic moment is present in a physical system that has a size between that of a neutron and a solenoid, the unconstrained description may be appropriate while still allowing an observation of the motion of the magnetic moment. Even, this scenario is plagued with an additional difficulty. For a finite system of charge and current distribution the electric and magnetic fields must approach zero at large distances from the charges and currents. Consider a charge and current loop that scatter from each other. When the charge and current loop are far apart the electromagnetic field momentum tends to zero. The total mechanical momentum must thus be identical for the final and initial state and Newton's third law holds. These statements imply that there is no difference between the constrained and unconstrained approach as far as momentum exchange is concerned. This statement may appear to be at odds with our above argumentation, but is not. The result of the imbalance of forces, and the violation of Newton's third law during the interaction at close proximity of the two interacting parts of the system, is a displacement for the final states, not a momentum exchange. This is not a general property, but can be shown in the impulse approximation for our unconstrained (equation (20)) and constrained force (equation (35)) by integrating the force over time for a straight path. Effects that depend on the differential cross section, such as the Sherman function for the Mott-Schwinger effect, are thus not expected to depend on the effective constraint in such a classical treatment.

Although, testing of unconstrained forces for A-B systems appears to be out of reach, a test of the Feynman paradox may be possible with current technology. Such a test would provide the first demonstration of the violation of Newton's third law (as it applies to the instantaneous conservation of mechanical momentum). Consider two electrons that are cross fired at each other. The capability to generate femtosecond electron pulses from nanoscale sources [30-32] gives control over the initial conditions of the trajectories that these electrons will follow. For electrons of about 1 keV energy the point of closest approach is on the order of microns. The capability to influence the motion of electrons in flight with a focused, pulsed laser may provide a means to make a "movie" of the electrons' trajectory. If momentum is stored in the electromagnetic field as Feynman states then controlling and monitoring both electron trajectories should reveal this behavior. Even with current technology, this is a major experimental challenge and perhaps explains why the Feynman paradox has never been demonstrated.

## Acknowledgements

This work was supported by the National Science Foundation under grant no. 0969506.